# Axial Profiling of Interferometric Scattering Enables an Accurate Determination of Nanoparticle Size†

Kateřina Žambochová,[a,b] Il-Buem Lee,[a] Jin-Sung Park,[a] Seok-Cheol Hong*[a,c] and Minhaeng Cho*[a,d]

Interferometric scattering (iSCAT) microscopy has undergone significant development in recent years. It is a promising technique for imaging and tracking nanoscopic label-free objects with nanometer localization precision. The current iSCAT-based photometry technique allows quantitative estimation for the size of a nanoparticle by measuring iSCAT contrast and has been successfully applied to nano-objects smaller than the Rayleigh scattering limit. Here we provide an alternative method that overcomes such size limitations. We take into account the axial variation of iSCAT contrast and utilize a vectorial point spread function model to uncover the position of a scattering dipole and, consequently, the size of the scatterer, which is not limited to the Rayleigh limit. We found that our technique accurately measures the size of spherical dielectric nanoparticles in a purely optical and non-contact way. We also tested fluorescent nanodiamonds (fND) and obtained a reasonable estimate for the size of fND particles. Together with fluorescence measurement from fND, we observed a correlation between the fluorescent signal and the size of fND. Our results showed that the axial pattern of iSCAT contrast provides sufficient information for the size of spherical particles. Our method enables us to measure the size of nanoparticles from tens of nanometers and beyond the Rayleigh limit with nanometer precision, making a versatile all-optical nanometric technique.

## Introduction

Interferometric scattering (iSCAT) microscopy is a label-free optical method for detecting, localizing, and tracking nanoparticles,[1-5] measuring the orientation of nanorods,[6] and imaging cellular components[7-9] with high spatial and temporal resolutions. The broad spectrum of iSCAT applications is possible due to its distinctive features like stable homodyne detection, high sensitivity, and unlimited observation time.[10-13] In this paper, we explore an application of iSCAT microscopy to nanoparticle (NP) size characterization.

The iSCAT signal from an NP small enough to ignore the pure scattering term is known to be proportional to the volume of the particle.[1,2] It was reported that the volume of a nanometric object such as NPs[14] and proteins[1,2] could be measured just from the iSCAT contrast. To truly quantify the size of an NP from scattering, however, we need to consider several factors contributing to the iSCAT signal, such as material properties of the scatterer, illumination wavelength, optical system properties (for example, numerical aperture (NA), size of beam waist, and polarization), refractive index mismatch, and optical aberration.[15-17] Another obstacle comes when the size of the NP exceeds the limit of Rayleigh scattering and the scattering becomes asymmetric.

Here, we consider small scatterers far below the Abbe diffraction limit but not necessarily smaller than the Rayleigh scattering limit. Although we cannot obtain images of the actual shape of an NP, we can determine the vertical position of its scattering dipole with the aid of rigorous theoretical calculation. Placing the source of this dipole, i.e., NP, on the cover glass surface, we demonstrate an accurate measurement of the height of this dipole from the cover glass, which, as we confirmed, quantitatively correlates with the size of the NP. In this way, we can measure the size of NPs.

Our method exploits the variation of the iSCAT signal over an extended range of vertical ($z$) focus scanning and utilizes a theoretical model to fit the experimental data. Successful reconstruction of experimental data by the theoretical model with suitable parameters enables us to extract quantitative information about individual NPs with nanometric accuracy, making it stand out amongst NP characterization techniques such as AFM or TEM for its all-optical approach and compatibility with the aqueous environment. As an optical method, it is also capable of simultaneous 3D tracking and bioimaging. Amongst optical techniques, through-focus scanning optical microscopy was introduced to determine the size of NPs via a calibration-based method.[18] Our approach is different from other optical methods

[a.] Center for Molecular Spectroscopy and Dynamics, Institute for Basic Science (IBS), Seoul 02841, Republic of Korea.
[b.] Department of Natural Sciences, Faculty of Biomedical Engineering, Czech Technical University in Prague, Kladno, 272 01, Czech Republic.
[c.] Department of Physics, Korea University, Seoul 02841, Republic of Korea.
[d.] Department of Chemistry, Korea University, Seoul 02841, Republic of Korea.
*Email (S.-C. Hong): hongsc@korea.ac.kr
*Email (M. Cho): mcho@korea.ac.kr

† Electronic Supplementary Information (ESI) available: Derivation of the formulae for the vectorial point spread function (PSF) model based on the Mie theory and their application to iSCAT detection. PSF analysis via Root-Mean-Square-Error (RMSE) evaluation for the determination of particle size. Nanoparticle tracking analysis (NTA) for the determination of nanoparticle size. Transmission electron microscope (TEM) images of fND samples.
See DOI: 10.1039/x0xx00000x

because it accomplishes both optical imaging and model-based size measurement of individual NPs. Theoretical models similar to the one used in the present work have been used in various studies: to localize metallic NPs in a medium by scattering microscopy,[15] to study metallic NPs by photothermal imaging,[16,17] and to compare scattering properties of metallic and dielectric NPs.[19] However, those studies did not address the quantitative measurement of NP size. With our approach, we successfully measured the size of spherical NPs with nanometric accuracy and precision without direct contact.

Besides introducing a new approach to measure the size of NPs with the iSCAT signal, we also gain a better insight into scattering signals from high-refractive-index NPs such as fluorescent nanodiamonds (fNDs) and their fluorescence properties. Our study reveals a correlation between the iSCAT contrast and fluorescence intensity. The axial profiling of interferometric scattering proposed here would be a versatile technique in nanoscience for NP size determination, together with the distinct advantages of iSCAT.

## Materials and methods

### Fluorescence combined iSCAT microscope (F-iSCAT)

The iSCAT setup and its fluorescence extension were already described in detail in the previous reports.[6,10] Fig. S1a shows the simplified schematics of F-iSCAT. The light (532 nm) from the laser (OBIS-FP-532LX, Coherent, USA) is steered in both $X$ and $Y$ directions using the acousto-optic deflector (AOD, DTS-XY400-532, AA Optoelectronics Ltd, France). It is sent through the $4f$ telecentric lens system (L1 and L2, AC254-500-A, Thorlabs, USA), the polarizing beam splitter (PBS, CCM1-PBS25-532, Thorlabs, USA), and the quarter-wave plate (QWP, WPQ10M-532, Thorlabs, USA) onto the back-focal plane of the objective lens (O, PLAPON60XO, oil-immersion, NA = 1.42, Olympus, Japan). The sample is placed on the high precision $XYZ$-piezo stage (P-545.3C8S, Physik Instrumente, Germany). The back-scattered light is then reflected by the dichroic mirror (DM, Di03-R532-t1-25´36, Semrock, USA) and projected through a lens (TL1, same as L1) on the sCMOS (pco-edge 4.2, PCO, Germany). The emitted fluorescence light is projected through a lens (TL2, same as L1) on the EMCCD (iXon897, Andor, UK). The stray light and leakage of the excitation beam are blocked by a notch filter (F1, NF03-532E-25, Semrock, USA) and an emission filter (F2, FF01-709/167-25, Semrock, USA).

The iSCAT and fluorescence signals are detected simultaneously. The iSCAT signal containing the back-scattered signal from an NP was recorded over a range (3 ~ 6 μm) of focal depth centered at the vertical position of the NP by moving the sample stage along the $z$-axis (Fig. S1b). Laser power was adjusted according to the size of the NP to avoid the saturation of both scattering and fluorescence intensities. All other parameters were set to be the same for all NP samples.

### Nanoparticles and their characteristics

We used a 40-nm polystyrene (PS) bead sample (FluoSpheres™ Carboxylate-Modified Microspheres, fluorescent (540/560), F8792 (dia.: 0.047±0.0055 μm) ThermoFisher Scientific, USA) and two latex bead samples (Amidine latex bead, 4%, '0.1 μm', A37313 (dia.: 0.10±0.009 μm), Life Technologies, USA; Aldehyde/Sulfate latex bead, 4%, '0.1 μm', A37287 (dia.: 0.12±0.009 μm), Life Technologies, USA). Hereafter, we mention their nominal radius, $R_{nom}$ = 20, 50, and 60 nm for the PS bead and two latex beads, respectively.

We also used a commercial fND sample (900172, Sigma-Aldrich, USA), which has one to four nitrogen vacancy (NV$^-$) centers per particle. We measured the size of the fND particle with nanoparticle tracking analysis (NTA) (NanoSight LM10, Malvern Panalytical, UK) or transmission electron microscopy (TECNAI G2, FEI, USA).

### Sample preparation for iSCAT measurement

All NP samples were sufficiently diluted from a stock solution and then sonicated for homogeneous mixing so that individual particles were sparsely and uniformly distributed for precise localization at the single-particle level when spread out on the glass surface. For PS and latex beads, the diluted samples were pipetted into the sample chamber (μ-Slide I Luer, 80167, ibidi, Germany), where the negatively charged beads were bound to the glass surface. Similarly, we pipetted the diluted fND sample into the sample chamber and washed out suspending particles to prevent overloading and forming clusters.

### Measurement of the iSCAT contrast and fluorescence from nanoparticles.

The iSCAT signal results from the interference between the scattered field from the sample and the reference field reflected from the cover glass surface. The total iSCAT intensity at the detector is given by

$$I = |E_R|^2 + |E_S|^2 + 2|E_R||E_S|\cos\Phi, \quad (1)$$

where $E_R$, $E_S$, and $\Phi$ are the reference field, the scattered field, and the relative phase between the two, respectively. The image-acquisition-based iSCAT microscope enables us to detect spatial intensity distribution. For an isolated NP, its spatial intensity distribution is known as the point spread function (PSF), which describes a diffraction-limited spot with radially symmetric fringes around the particle. The PSF images of a particle at different focal depths shown in Fig. S1c (i-iv) are distinctive because the scattered field with a spherically propagating wavefront and the reference field with the planar wavefront interfere with each other with a different relative phase and the spherical aberration by the optical system under non-ideal conditions adds additional complexity to the images. The stacked horizontal intensity profile (SHIP) of the iSCAT signal clearly shows how the intensity profile evolves with the vertical position of a scatterer relative to the focal plane. A particle to be analyzed defines a region of interest (ROI) whose dimension is 7×7 pixels and whose center coincides with the position of the particle. The

iSCAT image of an NP at each $z$-position was fitted with a 2D Gaussian function as follows:

$$\text{PSF}(x,y) = A \cdot \exp\left(-\left(\frac{(x-x_0)^2}{2\sigma_x^2} + \frac{(y-y_0)^2}{2\sigma_y^2}\right)\right) + B, \quad (2)$$

where $A$ and $B$ are the fitting parameters, $x_0$ and $y_0$ are the coordinates of the center, and $\sigma_x$ and $\sigma_y$ are the widths of the 2D Gaussian function. If the center of the NP image ($x_c$, $y_c$) matches the center of the 2D Gaussian function ($x_0$, $y_0$), then the parameter $A$ is the amplitude, and $B$ refers to the intensity offset, i.e., background, in the image. If we apply this to eqn (1), we get $|E_S|^2 + 2|E_R||E_S|\cos\Phi = \text{PSF}(x_0, y_0) - B$ and $|E_R|^2 \cong B$. Then the signal minus the background of the image followed by normalization at the center is

$$I_{\text{exp}}(x_0, y_0) = \frac{A}{B} = (|E_S|^2 + 2|E_R||E_S|\cos\Phi)/|E_R|^2, \quad (3)$$

which is referred to as interferometric scattering contrast (iSCAT contrast).

The fluorescent signal from an fND particle was measured in the fluorescence detection channel of F-iSCAT. An fND particle appeared as a round spot in the channel, and the PSF of the spot was also described by the same eqn (2). Thus, we calculated the fluorescence contrast ($I_{\text{fl}}$) by taking the ratio of the amplitude ($A$) and background offset ($B$), that is, $I_{\text{fl}} = A/B$.

## Result and Discussion

**Point spread function model for iSCAT imaging**

To understand the scattering from NPs and extract the information about them, we need to use an appropriate model to describe our system. A successful theoretical model should reproduce the PSF of an NP. Such a PSF model could be derived using either a vectorial or scalar approach. Scalar models derived from the diffraction theory of light use the approximation of the Fresnel-Kirchhoff integral to describe the propagation of a spherical wave through an aperture.[20] The scalar diffraction model by Gibson and Lanni (G-L) is computationally simpler and practically convenient as it directly introduces the experimental conditions as input parameters.[21,22] It calculates the imaging aberration from the optic path difference (OPD) between the design and experimental conditions of the layers between the objective and the sample.[23]

On the contrary, the vectorial model by Richards and Wolf (R-W) is rather complicated, but it provides an accurate ray tracing method for a radiating dipole in a focused beam.[24,25] Their vectorial approach is based on Maxwell's equations and calculates the electromagnetic field vectors,[20,26,27] the $x$, $y$, and $z$ components of which need to satisfy the corresponding wave equation.[27] The R-W's model was later reformulated by Török and Varga[28] (T-V) for more general use when the electromagnetic waves are focused through a stratified medium with mismatched refractive indices. Haeberlé[21,29] then demonstrated that such vectorial models could be used together with the G-L OPD, which provides an accurate and convenient way to model the PSF for optical microscopy. Therefore, we chose to simulate a fitting function for our experimental data with the T-V model using the G-L expression for OPD as a phase term.

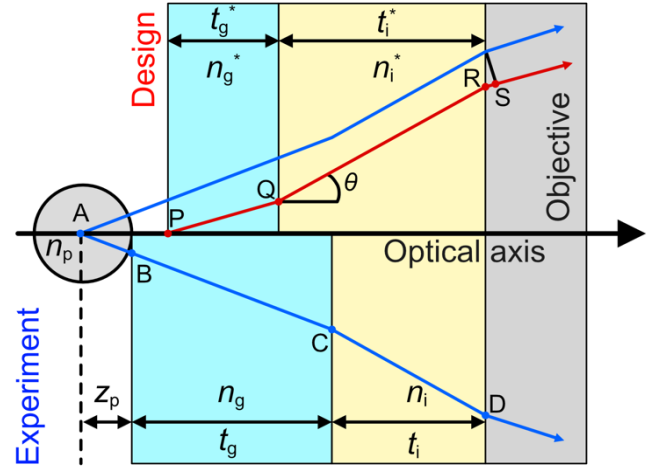

Fig. 1 Sample configuration and optical paths of ray ABCD under experimental conditions and PQRS under design conditions. $n$'s ($n_i$, $n_i^*$, $n_g$, $n_g^*$, and $n_s$) and $t$'s ($t_i$, $t_i^*$, $t_g$, and $t_g^*$) are the refractive indexes and thicknesses of various layers, respectively. The subscripts i, g, and s designate the immersion oil, cover glass, and sample, respectively. $z_p$ is the height of the induced dipole. $n_p$ is the refractive index of NP. Asterisk (*) is used to mark the design parameters.

The electric field at the observation space (focal region over the detector) can be expressed in the form of the R-W integral with the additional OPD term as follows:[30,31]

$$\vec{E}(\vec{r}) = -\frac{ik}{2\pi}\int_0^{2\pi}\int_0^{\theta_{d,\max}} \vec{E}_\infty(\theta_d,\phi)e^{ik\vec{s}\cdot\vec{r}}e^{ik\Lambda_{\text{OPD}}}f(\theta_d)\sin\theta_d\, d\theta_d\, d\phi, \quad (4)$$

where $\vec{E}_\infty(\theta_d,\phi)$ is the electric strength vector at the pupil of the imaging lens (far away from the focus of the lens, hence called 'far-field'), $\vec{s}$ the unit vector of field propagation, $\vec{r}$ the vector of the observation point, $\Lambda_{\text{OPD}}$ the optic path difference responsible for spherical aberration, $f(\theta_d)$ the apodization factor, $\theta_d$ the zenith angle of the ray focused by the imaging lens, and $\theta_{d,\max}$ the semi-aperture angle of $\theta_d$. $k$ is the wavenumber of the ray in the observation space, which is the same as the value in a vacuum. $\phi$ is the azimuthal angle in the optical system and defines the azimuthal direction of a ray. $\theta$ is the angle of a ray made with respect to the optic axis, and here it defines the polar direction of a ray in the sample space (immersion oil), which is related to $\theta_d$ by the following equation: $n_i \sin\theta = Mn_a \sin\theta_d$, where $M$ is the magnification of the imaging system. We calculate the reference ($E_R$) and scattered ($E_S$) fields at the focal region of the imaging lens (over the detector) from the respective 'far-field' electric strength vectors using eqn (4).

To derive $\vec{E}_\infty(\theta_d,\phi)$ for reference and scattered fields, we start with a plane wave with linear polarization ($\vec{E}_i$) and trace the vectorial components of the reference and scattered electric fields by the generalized Jones matrix formalism, as illustrated in Fig. S2. The strength vector of the 'far-field' reference field can be expressed as follows:

$$\vec{E}_{R,\infty}(\theta_d,\phi) = C_R \cdot i(r_p + r_s)\begin{bmatrix}-(1-\cos\theta_d)\cos\phi\sin\phi \\ \cos^2\phi + \sin^2\phi\cos\theta_d\end{bmatrix} \quad (5)$$

where $C_R$ is just an overall numerical factor, $r_p$ and $r_s$ are the Fresnel reflection coefficients at the glass-water interface and the incident field ($\vec{E}_i$) is linearly polarized along the $x$-axis.

The strength vector of the scattered field by an NP is usually described by the Rayleigh scattering theory, which is only valid for particles smaller than the Rayleigh scattering limit. Thus, we consider the Mie theory[32] instead, for accurate modeling of the scattered field from NPs whose size is beyond the Rayleigh regime. The strength vector of the 'far-field' scattered field can be expressed as follows:

$$\vec{E}_{S,\infty}(\theta_d,\phi) = C_S \cdot i(S_2 t_p + S_1 t_s) \begin{bmatrix} -(1-\cos\theta_d)\cos\phi\sin\phi \\ \cos^2\phi + \sin^2\phi\cos\theta_d \end{bmatrix} \quad (6)$$

where $C_S$ is the overall factor of the scattered field, which includes the collection efficiency $\eta$, $t_p$ and $t_s$ are the Fresnel transmission coefficients at the glass-water interface, and $S_1$ and $S_2$ are functions of $\theta$ found in the scattering field components from a spherical particle obtained by the Mie theory[32].

The iSCAT contrast depends on the phase difference between $E_R$ and $E_S$, which is determined by the size and geometry of a scattering object, illumination wavelength, refractive index mismatch in the optical path, and NA.[15,16,19] The phase term, according to the G-L model, accounts for the aberration caused by index mismatch and finite thickness of multiple layers along the optical path.[22] The aberration can be described by considering the OPD ($\Lambda$) between the real (experimental) and ideal (design) beam paths ( $\Lambda = [ABCD] - [PQRS]$ , where $[ABCD]$ and $[PQRS]$ are the pathlengths of real and ideal paths, respectively), as illustrated in Fig. 1.[22,23]

In the axial scan, the sample stage moves while the objective lens stays stationary, which results in changing the distance between the stage and the objective and therefore altering the thickness of the immersion oil layer. In that case, $t_i$ should be expressed in terms of other parameters as[20,23]

$$t_i = n_i \left( \frac{-z_f}{n_i} + \frac{t_g^*}{n_g^*} - \frac{t_g}{n_g} + \frac{t_i^*}{n_i^*} - \frac{z_p}{n_s} \right) \quad (7)$$

where $z_f = z - z_p$ is the displacement of the focal plane from the particle, meaning that the particle is best focused at $z_f = 0$.[20] Thus, we position the minimum point of the experimental axial profile at $z_f = 0$ as shown in Fig. 2.

To express the OPD in terms of the parameters of the relevant layers (immersion oil, cover glass, sample), we resort to Snell's law of refraction and find that[20]

$$\Lambda(\theta, z_f, z_p, \tau) = z_p\sqrt{n_s^2 - n_i^2\sin^2\theta}$$
$$+ t_i\sqrt{n_i^2 - n_i^2\sin^2\theta} - t_i^*\sqrt{n_{i*}^2 - n_i^2\sin^2\theta} \quad (8)$$
$$+ t_g\sqrt{n_g^2 - n_i^2\sin^2\theta} - t_g^*\sqrt{n_{g*}^2 - n_i^2\sin^2\theta}$$

where $\tau = (n_i, n_i^*, n_g, n_g^*, n_s, t_i^*, t_g, t_g^*)$ and $t_i$ is given by eqn (7).

The OPD of the reference field ($\Lambda_{\text{OPD,Ref}}$) is given by those of the normally incident and reflected rays with respect to ideal rays along the same path:

$$\Lambda_{\text{OPD,Ref}} = 2(n_i t_i - n_i^* t_i^* + n_g t_g - n_g^* t_g^*). \quad (9)$$

The scattered field depends on the height of the dipole (center position of the spherical scatterer) and thus the OPD of the field ($\Lambda_{\text{OPD,Scat}}$) can be calculated using the ray geometry considered by the G-L formalism. The OPD of the scattered light is given as:

$$\Lambda_{\text{OPD,Scat}} = \Lambda(\theta, z_f, z_p, \tau) + n_s z_p + n_i t_i - n_i^* t_i^* + n_g t_g - n_g^* t_g^*. \quad (10)$$

The reference (eqn (5)) and scattered (eqn (6)) fields are inserted in the R-W integral (eqn (4)) with the aforementioned OPD terms and the integral is evaluated over $\phi$, which leads to the simplified form (eqn (S11-S13)). Then, the iSCAT contrast is computed by subtracting the background from the signal and normalizing the difference with the background as:

$$I_{\text{sim}} = \frac{|\vec{E}_S|^2 + 2|\vec{E}_R||\vec{E}_S|\cos\Phi}{|\vec{E}_R|^2}. \quad (11)$$

To reproduce experimental results by computation, we calculated the reference field by reflection and the scattered field from a dielectric particle as a function of the relative position of the focal plane ($z_f$) and visualized the SHIP of electric field amplitude for the reference, scattering, and iSCAT ($I_{\text{sim}}$) in Figs. 2b-d.

As shown in Figs. 2b and c, the reference and scattered fields propagate with the planar and quasi-spherical wavefronts, respectively. Moreover, the SHIP images and their axial cuts

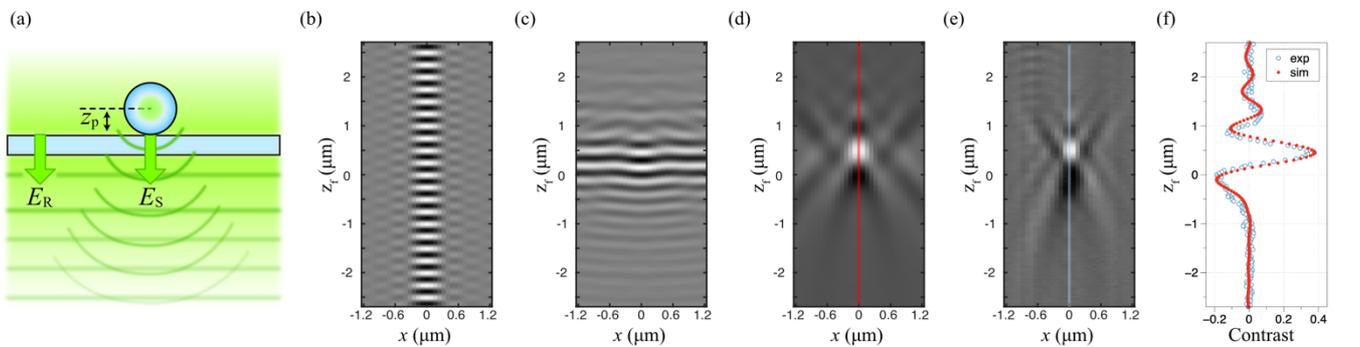

Fig. 2 (a) Propagation of the scattered field ($E_S$) from a nanoparticle and the reference field ($E_R$) from the sample-surface interface. SHIPs of the PSF of the reference field (b), the scattered field (c) and the iSCAT signal (d) are displayed as a function of focal position ($z_f$). (e) SHIP of the PSF experimentally acquired for a latex NP (diameter = 120 nm). (f) Axial intensity profile of the center of a nanobead (experimental: blue circles; contrast along the blue line in (e); simulated: red dots)). The minimum is placed at $z_f$ =0.

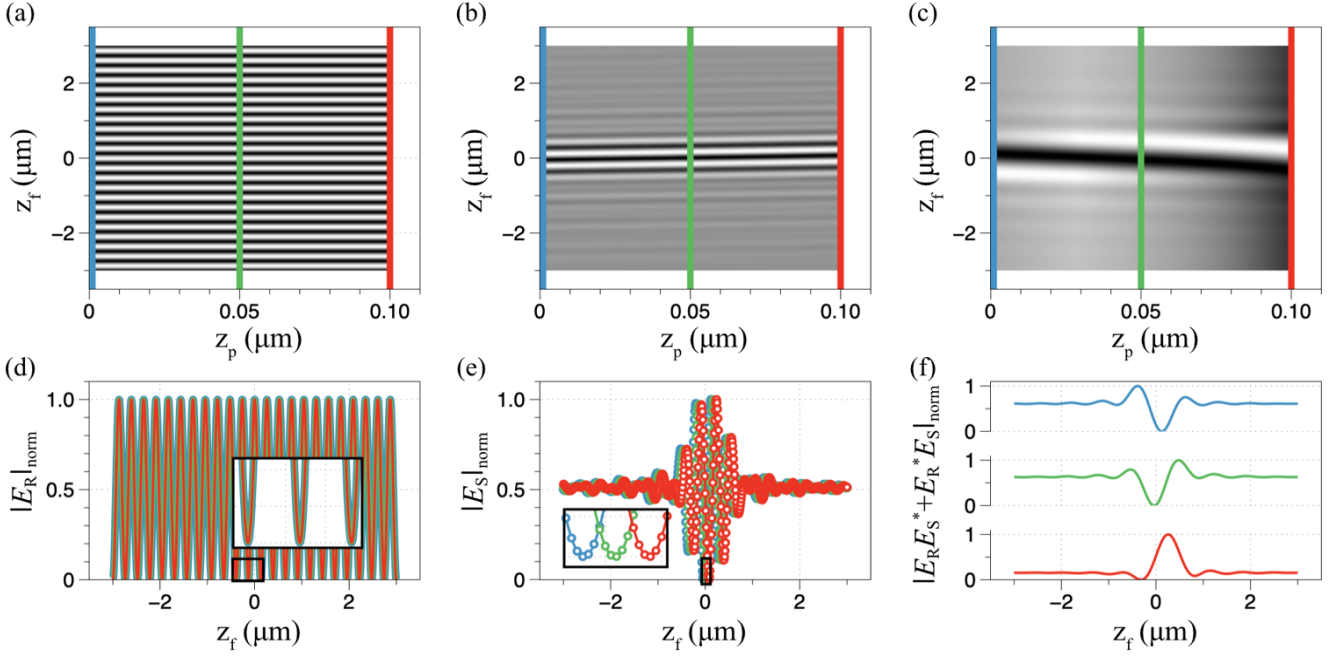

**Fig. 3** Stacked axial contrast profile as a function of $z_p$. (a) $|E_R|$, (b) $|E_S|$ and (c) $|E_R E_S^* + E_R^* E_S|$ calculated in the range of $z_p = 0 \sim 100$ nm with $n_i = 1.52$. All calculated values are normalized and scaled to fit to [0,1] and displayed in gray scale (hence, labelled as $|E_R|_{norm}$, $|E_S|_{norm}$, and $|E_R E_S^* + E_R^* E_S|_{norm}$ in (d-f), respectively). The blue, green and red colored lines indicate $z_p = 0$, 50, and 100 nm. (d-f) Cross-sections of the reference field (d), the scattered field (e), and the interference term (f). Insets in (d) and (e) show the phase variation of the fields and highlight the sensitivity of the fields to different values of $z_p$. The axial intensity profile by the interference term exhibits more pronounced variation with $z_p$, accounting for sensitive detection of the size of nano-objects by iSCAT. In this figure, $z_f$ is the computational input parameter before the offset ($\delta z$) is adjusted.

(axial variation of iSCAT signal at the center of an NP) from the actual measurement and numerical computation look almost identical, as illustrated in Figs. 2d and e. Our analysis showed that we could produce simulation results highly similar to experimental observations with only a specific set of parameters, indicating that our approach is robust and reliable (Fig. 2f).

**Strong dependence of dipole's iSCAT signal on its height as the basis for accurate measure of the size of a dielectric sphere**

The interferometric signal between reference and scattered fields critically depends on the OPD variation, which originates from the axial location of the NP relative to the reference interface. Thus, the location of the scattering source would affect the shape of the wavefront and thus the spatial variation of the phase. It is well known that the incident field induces and drives the dipole moment in a dielectric nanoparticle. Therefore, we simulated the propagation of fields using the PSF model discussed above in order to understand how the height of the dipole, equivalently, the radius of the spherical NP affects the phase variation or the whole signal profiles. The numerical results as a function of the height of the dipole ($z_p$) are displayed in Figs. 3a-c, together with the axial profiles for three representative values of $z_p$ (0, 50, and 100 nm) shown in Fig. 3d-f. As shown in Figs. 3a and d, $E_R$ has no dependence on the size of an NP as expected. On the contrary, the modulation of the scattered field slightly shifts along the axial direction ($z_f$) as $z_p$ changes as shown in Fig. 3e (see inset). This phase shift is responsible for the axial variation of the iSCAT signal shown in Fig. 3f as well as in Figs. S3 and S4, which show that a small change in $n_i$ influences the phase variation of $E_S$ significantly, indicating that $n_i$ needs to be precisely determined for accurate measurement of $z_p$.

Thus, a small change in $z_p$ results in a notable change in the axial profile, which warrants a reliable and unambiguous analysis of experimental data. The scattering signal itself varies in amplitude, but due to its interference with the reference field, the change in the shape and amplitude of the interferometric signal becomes more pronounced. Therefore, the axial profile of the iSCAT signal significantly enhances the sensitivity to particle size. It becomes clear that our approach based on the quantitative fitting analysis of the axial pattern of the iSCAT signal can be an incisive tool for distinguishing NPs with different sizes.

**Reconstruction of axial PSF variation**

The realistic model and accurate fitting of experimental data can be achieved only with fitting parameters close to reality. In the computation of PSF, the wavelength of illumination and the NA used were 532 nm and 1.35, respectively. The refractive indexes and thicknesses used were: $n_s = 1.33$ (for water), $n_p = 1.59$ (for polystyrene and latex nanoparticles) or 2.4 (for nanodiamond particles), $n_g = n_g^* = 1.52$ (for cover glass), $n_i^* = 1.52$ (for immersion oil), $t_g = t_g^* = 170$ μm (for cover glass), and $t_i^* = 100$ μm (for immersion oil). Symbols with and without asterisk represent parameters in real (experimental) and ideal (design) conditions as used in the G-L model (eqns (7), (8)).[23] The $t_i$, 'real' parameter of oil thickness, is a parameter to be adjusted in the G-L model.

We varied $n_i$ and $z_p$ to describe the spherical aberration in the axial profile. Numerical values of these fitting parameters were determined by the root mean square error (RMSE) evaluation. First, we define the parameter space for $n_i$ and $z_p$ to be examined. For each set of $n_i$ and $z_p$ values, we generate the PSF (equivalently, $I_{exp}$) at the focal space of the imaging lens (equivalently, on the detector surface) with various depths of the focal plane of the objective (i.e., $z_f$ in eqn (7) or $z$ in the definition of $z_f$) from the R-W integral in eqns (S7,8,11-13). Typically, we vary $z$ within a few μm from $z = 0$. Since we are only interested in its center value, $I_{exp}(x_0, y_0)$, we can get the axial profile of the center of the PSF by setting $r = 0$ in the R-W integral. This simulated PSF ($I_{sim}$) is likely to be off with respect to $I_{exp}$ due to the non-ideal experimental parameters (causing spherical aberration) and the finite size of scatterers, thus we shift $I_{sim}$ by $\delta z$ to find the optimal $I_{sim}$ with the lowest RMSE for the given $n_i$ and $z_p$ by compensating for a potential offset (the compensating offset giving the least RMSE is denoted as $\delta z^*$). By sequentially varying $\delta z$ (distance translated for $I_{sim}$ with respect to $I_{exp}$ as illustrated in Fig. S5a) with a fine increment within a reasonable range, one would find the least RMSE with the best $\delta z^*$ (Fig. S5b) and obtain $I_{sim}$ that best matches $I_{exp}$ under the given set of $n_i$ and $z_p$. This least RMSE is called the representative RMSE value for the given $n_i$ and $z_p$.

**Axial profiling of the PSF center enables accurate determination of nanoparticle size.**

We carried out the RMSE estimation for axial intensity profiles generated with various values of $n_i$ and $z_p$ as shown in Fig. S6a. From the minimization of RMSE, we could determine the values of $n_i$ and $z_p$ as shown in Fig. S6b. To validate our approach, we tested polystyrene (PS) and latex beads because they are highly uniform in size and shape (spherical), not to mention that they are homogeneous. The size uniformity of beads was confirmed by our NTA and TEM measurements shown in Fig. S7a-c.

Using the PSF model and adjusting the parameters carefully, we were able to make the computational results nearly identical to the corresponding experimental data: the experimental and theoretical SHIP images are remarkably similar for different-sized beads as shown in Fig. S8. Interestingly, the results from those NPs showed that $z_p$ determined by our method is in excellent agreement with the expected radius of those NPs. From our model fitting, we obtained the sizes of PS and latex beads, which closely match their $R_{nom}$'s in Fig. 4a: (i) for 20-nm PS beads, $z_p \sim 15$ ($\pm$ 9.5) nm; for this PS bead, the NTA method failed to give reasonable results because it is substantially smaller than the size limit for reliable measurements by NTA (dia. $\sim$ 60 nm with 642 nm excitation[33]) but TEM provided R $\sim$ 23($\pm$ 2.8) nm, consistent with the result from our PSF modeling; (ii) for 50-nm latex beads, we got $z_p \sim 40$ ($\pm$ 7.8) nm while NTA and TEM gave R $\sim$ 50 ($\pm$ 18) nm and $\sim$ 50 ($\pm$ 3) nm, respectively, both supporting the result from PSF modeling; (iii) for 60-nm latex beads, we got $z_p \sim 57$ ($\pm$ 2.7) nm while NTA and TEM gave R $\sim$ 55 ($\pm$ 21) nm and $\sim$ 60 ($\pm$ 3) nm, respectively, supporting the result from PSF modeling again. As shown here, TEM is a reliable and accurate tool to measure the size of NP by direct visualization, but the limited accessibility of TEM and the low throughput and technical difficulty of the technique hampers easy and wide applications to samples in the condensed phases. This dipole-modeling-based result is consistent with our view that the dipole of those scatterers is located near the centre of the particles and the particles are in good contact with the interface.

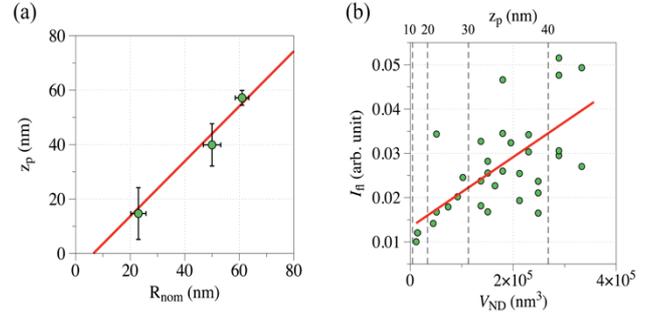

**Fig. 4** Representative results of particle size determination by PSF analysis. (a) Accuracy of the axial profile method to determine the size of spherical nanoparticles (PS and latex bead). $z_p$, radius of bead acquired by the axial profile method, is not only linearly correlated with but also nearly identical to 'Nominal radius ($R_{nom}$)' provided by manufacturers. Linear fit $y = 1.01 \cdot x - 6.6$ with $R^2 = 0.97$. (b) Correlation of the fluorescence intensity ($I_{fl}$) and volume ($V_{ND}$) of '40 nm' fND together with a linear fit of the data. For a better idea about the fND size, the upper x-axis shows the value of $z_p$ determined by iSCAT. Several values of $z_p$ are indicated by vertical dashed lines. Correlation coefficient: $\mathcal{R}(V_{fl}, V_{ND})_{40} = 0.57$.

Next, we estimated the size of fND using the PSF model. fNDs have drawn much attention for their remarkable properties and wide applications in biological imaging and sensing.[34,35] We generated theoretical SHIPs of fNDs with carefully adjusted parameters, which are in good agreement with the experimental SHIPs as shown in Fig. S9. The measured sizes ($z_p$) of fNDs were broadly distributed, as shown in Fig. 4b, consistent with the TEM result shown in Fig. S7d. It, however, turned out that the size of an fND estimated by our approach was correlated with the fluorescence intensity of the fND. We shall discuss this issue in the next section.

Although NTA is a well-known and popular tool for size determination of NPs, it has several technical pitfalls[36], which our method could overcome. First, the size distribution measured by NTA was considerably broader than that acquired by our axial profile-fitting method. It is likely because our single-particle method enables us to choose individual NPs and avoid clusters, significantly larger NPs, and impurities in the sample, which is intrinsically impossible with the NTA method. Second, the NTA is less straightforward in measuring the size of small NPs because instrumental settings and user inputs are critical for such NPs. Although the NTA could be of some use for measuring NPs of size below 60 nm, the uncertainty would be large, as exhibited in Fig. S7. In contrast, the sensitivity of our method is high, so our approach enables accurate measurements of the size of PS beads as small as 40 nm and that of fNDs down to 30 nm in diameter (Fig. 4).

**Correlation of fND volume and fluorescence intensity**

fNDs are fluorescent owing to the negatively charged NV$^-$ centres in their diamond lattice. The fluorescence intensity $I_{fl}$ is proportional to the number of NV$^-$ centres in a given fND. From

$N_{NV} = c_{NV} \cdot V_{ND}$ ($c_{NV}$: concentration of NV⁻ centres, $V_{ND}$: volume of fND), the fluorescence intensity should also increase linearly with the volume of fND. Regarding the correlation between them, we shall consider the following degrading factors: (i) inhomogeneous incorporation of NV⁻ centres throughout nanodiamond lattice, (ii) the inaccurate assumption that the fNDs have a spherical shape, in fact, our fNDs appear significantly jagged as confirmed by TEM (Fig. S7d), (iii) weaker fluorescence emission by NV⁻ centres placed near the particle surface than in the centre of the particle[37], and (iv) batch-to-batch variation of the number of NV⁻ centres in the fabrication process.

Fig. 4b shows the relationship between the volume of fND, $V_{ND}$, calculated from its radius, $z_p$, determined by iSCAT, and the fluorescence intensity $I_{fl}$ of fND. To describe the correlation between the fluorescence intensity and fND volume, we used the Pearson correlation coefficient and obtained the correlation $\mathcal{R}(V_{fl}, V_{ND})_{40} = 0.57$, which signifies only a moderate correlation. There are several particles that deviate considerably from this correlation. The main reasons for scattered points would be, as assumed above, non-uniform incorporation of NV⁻ centres in the diamond lattice and the non-spherical shape of the nanoparticles. This implies that the size of fNDs cannot be deduced accurately from their fluorescence intensity alone. On the other hand, we found that the distributions of fND size obtained by different methods (PSF, TEM, NTA) were well overlapped, supporting that the value of $z_p$ (or the size of NP) acquired by the PSF model analysis is valid. From this, we suggest that the PSF analysis proposed here can even determine the size of NPs of non-spherical shape such as fNDs with reasonable accuracy.

## Conclusions

The iSCAT microscopy has evolved into a useful label-free optical technique that enables both imaging and tracking nanoscopic objects with high precision. Here, we demonstrated that it could be used to characterize the size of NPs all-optically even beyond the Rayleigh scattering regime. We developed a technique useful to measure the size of individual NPs by acquiring the axial variation of the iSCAT signal and fitting the theoretical PSF model to the axial profile of the iSCAT signal, which provides the information on the scattering dipole position, *i.e.*, size of NPs in the present case.

The theoretical model used to fit the experimental data is a modified Török and Varga's vectorial PSF theory. Our model accounts well for the factors that contribute to the iSCAT signal resulting from the interference between the scattered and reference fields. Our method turns out to be sensitive to small changes (within a few nanometers) in the size of NPs, which surpasses other optical methods such as NTA or DLS (Dynamic Light Scattering) in size sensitivity and single-particle characterization capability, not just ensemble size distribution. It also stands out amongst other size determination techniques for its instrumental simplicity and all-optical, contactless, nondestructive, and non-contaminating approach. The results presented here demonstrate that the axial profile-fitting method is a useful approach not only because it can be used to measure the size of NP covering a broader range from ~ 10 nm to several hundred nanometers, well-beyond Rayleigh scattering limit[2,14] but also because it can be applied to NPs made of various materials. All taken together, we anticipate that our technique would be useful for characterizing NPs and stratified media.

## Author contributions

S.-C.H. and M.C. conceived the idea. K.Z. and I.-B.L. performed the experiments. I.-B.L. wrote MATLAB programs for modeling and fitting. K.Z. analyzed the data and prepared figures. K.Z., S.-C.H., M.C, I.-B.L. and J.-S.P. wrote the paper.

## Conflicts of interest

There are no conflicts to declare.

## Acknowledgments

This work was supported by the Institute for Basic Science (IBS-R023-D1). This work was also partly supported by an NRF grant (2022R1A2B5B01002343) and the Global Research and Development Center Program (2018K1A4A3A01064272) through the NRF of Korea (S.-C.H.) and Czech Technical University foundation SGS17/201/OHK4/3T/17.

# SUPPLEMENTARY INFORMATION

## Axial Profiling of Interferometric Scattering Enables an Accurate Determination of Nanoparticle Size

Kateřina Žambochová,[a,b] Il-Buem Lee,[a] Jin-Sung Park,[a] Seok-Cheol Hong*[a,c] and Minhaeng Cho*[a,d]

Interferometric scattering (iSCAT) microscopy has undergone significant development in recent years. It is a promising technique for imaging and tracking nanoscopic label-free objects with nanometer localization precision. The current iSCAT-based photometry technique allows quantitative estimation for the size of a nanoparticle by measuring iSCAT contrast and has been successfully applied to nano-objects smaller than the Rayleigh scattering limit. Here we provide an alternative method that overcomes such size limitations. We take into account the axial variation of iSCAT contrast and utilize a vectorial point spread function model to uncover the position of a scattering dipole and, consequently, the size of the scatterer, which is not limited to the Rayleigh limit. We found that our technique accurately measures the size of spherical dielectric nanoparticles in a purely optical and non-contact way. We also tested fluorescent nanodiamonds (fND) and obtained a reasonable estimate for the size of fND particles. Together with fluorescence measurement from fND, we observed a correlation between the fluorescent signal and the size of fND. Our results showed that the axial pattern of iSCAT contrast provides sufficient information for the size of spherical particles. Our method enables us to measure the size of nanoparticles from tens of nanometers and beyond the Rayleigh limit with nanometer precision, making a versatile all-optical nanometric technique.

**SI contents**

[a.] Center for Molecular Spectroscopy and Dynamics, Institute for Basic Science (IBS), Seoul 02841, Republic of Korea.  
[b.] Department of Natural Sciences, Faculty of Biomedical Engineering, Czech Technical University in Prague, Kladno, 272 01, Czech Republic.  
[c.] Department of Physics, Korea University, Seoul 02841, Republic of Korea.  
[d.] Department of Chemistry, Korea University, Seoul 02841, Republic of Korea.  
*Email (S.-C. Hong): hongsc@korea.ac.kr  
*Email (M. Cho): mcho@korea.ac.kr

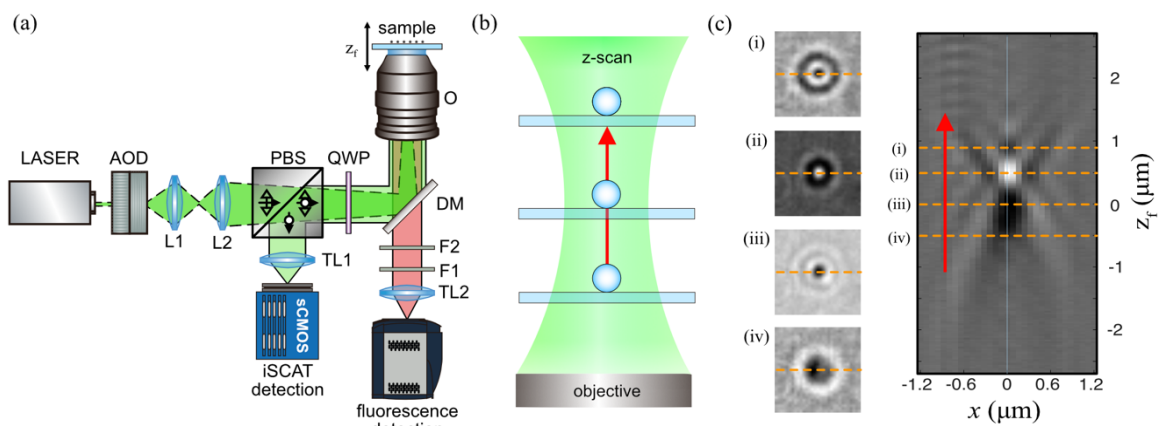

**Fig. S1** Schematic diagrams of the F-iSCAT setup and data acquisition by z-scanning. (a) Fluorescence combined iSCAT microscope. The outlined and dashed-outlined green beams mark the scattered and illuminating lights, respectively, and the red beam shows fluorescence from NPs. Optical elements in the setup are acousto-optic deflector (AOD), lenses (L1, L2, TL1, TL2), polarizing beam splitter (PBS), quarter-wave plate (QWP), objective (O), dichroic mirror (DM), notch filter (F1), and emission filter (F2). (b-c) Illustration of z-scanning and representative axial variation of iSCAT image by an NP (dia.: 120 nm). (b) Sample is scanned along the z-axis (red arrow) over a range of 5 μm around the focal plane by moving the sample stage. (c) Experimental SHIP image by stacking the cross-sectional intensity profile (guided on the orange dashed lines) along the diameter of the NP image (indicated by orange-dashed lines in (i-iv)).

**Vectorial point spread function modeling of nanoparticle image by iSCAT microscopy**

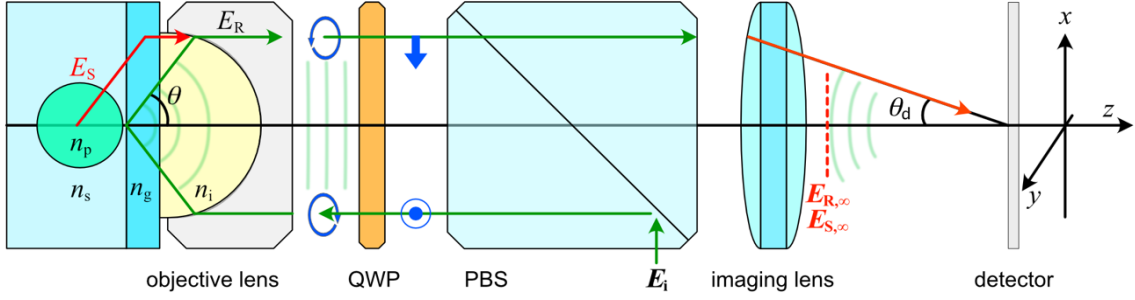

**Fig. S2** Illustration of the optical components of the iSCAT microscopy.

A generalized Jones matrix describes the effect of an optical element on ray propagation. Our system consists of a polarizing beam splitter (PBS), a quarter-wave plate (QWP), several lenses, and transmitting/reflecting interfaces for which we used the relevant Jones matrices.[1] The matrices were applied in order to trace the polarization state of light. We consider the incident beam that impinges on the PBS. By the PBS, the beam becomes 45° linearly polarized with respect to the optic axis of the subsequent QWP and circularly polarized at the back focal plane of the objective by the QWP. The beam shines at the sample through the objective lens, and the reflected and scattered fields propagate all the way to the dotted line after the imaging lens as illustrated in Figure S2, passing through the objective lens, QWP, PBS, and the imaging lens. The Richards-Wolf integral is then applied to those fields to find out the field at the focal region (detector).

**A. Derivation of the reference and scattered fields in the pupil region of the imaging lens**

<u>Reference field</u>: The reference field $E_R$ is the light reflected at the glass and water/sample interface. The reference field at the pupil of the imaging lens ($\vec{E}_{R,\infty}$) is given as:

$$\vec{E}_{R,\infty}(\theta_d, \phi) = \mathbb{R}_z^{-1}(\phi).\mathbb{L}(\theta_d).\mathbb{R}_z(\phi).\mathbb{P}(0°).\mathbb{W}(90°, 45°).\mathbb{R}_z^{-1}(\phi).\mathbb{F}_R.\mathbb{R}_z(\phi).\mathbb{W}(90°, 45°).\mathbb{P}(90°).\vec{E}_i \quad (S1)$$

where $\mathbb{P}(0°)$, $\mathbb{P}(90°)$, $\mathbb{W}(90°, 45°)$, $\mathbb{R}_z(\phi)$, $\mathbb{L}(\theta_d)$, and $\mathbb{F}_R$ are the Jones matrices for polarizing beam splitters for transmission and reflection, quarter-wave plate, coordinates rotation by $\phi$ about the optic axis for local p-/s-wave decomposition, imaging lens with the zenith angle $\theta_d$, and Fresnel field transformation by reflection, respectively. The incident field is given by $\vec{E}_i = (E_x, E_y) = (0,1)$.

$$\vec{E}_{R,\infty}(\theta_d, \phi) = C_R \cdot i(r_p + r_s)\begin{bmatrix}-(1-\cos\theta_d)\cos\phi\sin\phi \\ \cos^2\phi + \sin^2\phi\cos\theta_d\end{bmatrix} = \frac{1}{2}C_R \cdot i(r_p + r_s)\begin{bmatrix}-(1-\cos\theta_d)\sin 2\phi \\ (1+\cos\theta_d)+(1-\cos\theta_d)\cos 2\phi\end{bmatrix} \quad (S2)$$

where $r_p$ and $r_s$ are the Fresnel reflection coefficients for p- and s-polarized light and $C_R$ is just the overall numerical pre-factor.

<u>Scattered field</u>: In this work, we only deal with scattering by a spherical particle. The scattered field by a spherical scatterer ($\vec{E}_s$) takes a complex form, as shown below. The scattered field at the pupil of the imaging lens ($\vec{E}_{S,\infty}$) is expressed in terms of $\vec{E}_s$ defined in local spherical coordinates:

$$\vec{E}_{S,\infty}(\theta_d, \phi) = \mathbb{R}_z^{-1}(\phi).\mathbb{L}(\theta_d).\mathbb{R}_z(\phi).\mathbb{P}(0°).\mathbb{W}(90°, 45°).\mathbb{R}_z^{-1}(\phi).\mathbb{F}_T.\vec{E}_s \quad (S3)$$

where $\mathbb{F}_T$ is the Fresnel transmission matrix at the interface. If the incident field at the sample is circularly polarized, the scattered field is given by $\vec{E}_S = \hat{\theta} E_{S\theta} + i\hat{\phi} E_{S\phi}$. $\vec{E}_S$ is described in terms of the spherical coordinates $(r, \theta, \phi)$ defined in the sample chamber with the origin at the center of the spherical particle. $\theta$ is the angle of a ray made with respect to the optic axis, and thus, it defines the polar direction of a ray in the sample space (immersion oil). The scattered field from a spherical scatterer is acquired by the Mie-scattering theory; $E_{S\theta}$ and $E_{S\phi}$, the polar and azimuthal components of the scattered electric field, respectively, shall be expanded as a linear combination of $\pi_n = P_n^1/\sin\theta$ and $\tau_n = \dfrac{dP_n^1}{d\theta}$

$$E_{S\theta} = \frac{1}{\sqrt{2}} \frac{e^{ikr}}{-ikr} e^{i\phi} S_2(\theta, \phi) \propto e^{i\phi} \sum_n \frac{2n+1}{n(n+1)} (a_n \tau_n + b_n \pi_n) \tag{S4}$$

$$E_{S\phi} = \frac{1}{\sqrt{2}} \frac{e^{ikr}}{-ikr} e^{i\phi} S_1(\theta, \phi) \propto e^{i\phi} \sum_n \frac{2n+1}{n(n+1)} (a_n \pi_n + b_n \tau_n) \tag{S5}$$

where $S_1$ and $S_2$ are some functions defined in terms of $\pi_n$ and $\tau_n$, and $a_n$ and $b_n$ are the coefficients of $\pi_n$ and $\tau_n$.[2]

The final form of the scattered field is simplified as follows:

$$\vec{E}_{S,\infty}(\theta_d, \phi) = C_S \cdot i(S_2 t_p + S_1 t_s) \begin{bmatrix} -(1 - \cos\theta_d) \cos\phi \sin\phi \\ \cos^2\phi + \sin^2\phi \cos\theta_d \end{bmatrix} = \frac{1}{2} C_S \cdot i(S_2 t_p + S_1 t_s) \begin{bmatrix} -(1 - \cos\theta_d) \sin 2\phi \\ (1 + \cos\theta_d) + (1 - \cos\theta_d) \cos 2\phi \end{bmatrix} \tag{S6}$$

where $t_p$ and $t_s$ are Fresnel transmission coefficients at the glass and sample/water interface for p- and s-polarized light, and $C_S$ is the overall numerical pre-factor including $\eta = (1/\pi) \sin^{-1}(\min(NA/n_s, 1))$, collection efficiency of the objective lens along the angular extent of aperture.

**B. Electric fields at the focal region of the imaging lens derived from the Richards-Wolf integral**
The electric field at the focal region of an imaging lens is calculated by the R-W integral:

$$\vec{E}(\vec{r}) = -\frac{ik}{2\pi} \int_0^{\theta_{d,max}} \int_0^{2\pi} \vec{E}_\infty(\theta_d, \phi) e^{ik\vec{s}\cdot\vec{r}} e^{ik\Lambda_{OPD}} f(\theta_d) \sin\theta_d \, d\phi d\theta_d \tag{S7}$$

where the apodization function is given by $f(\theta_d) = \sqrt{\dfrac{n_i \cos\theta_d}{n_a \cos\theta}}$, $\vec{s} \cdot \vec{r} = (\sin\theta_d \cos\phi, \sin\theta_d \sin\phi, \cos\theta_d) \cdot (r\cos\phi_d, r\sin\phi_d, z_d) = r\sin\theta_d \cdot \cos(\phi - \phi_d) + z_d \cos\theta_d$, $\theta_{d,max}$ the semi-aperture angle of $\theta_d$, $k = \dfrac{2\pi n_a}{\lambda}$ the angular wavenumber of light, $n_a$ the refractive index of air, $r = \sqrt{x^2 + y^2}$ the distance from the focus to an observation point, and $\phi = \tan^{-1}(s_y/s_x)$ the azimuthal angle of an imaging ray. $\vec{E}_{R,\infty}(\theta_d, \phi)$ and $\vec{E}_{S,\infty}(\theta_d, \phi)$ are inserted in eqn (S7) in the place of $\vec{E}_\infty(\theta_d, \phi)$ to get $\vec{E}_R$ and $\vec{E}_S$, respectively. As a ray propagates through a series of optical systems, the (polar) angle made by the ray with respect to the optic axis changes. To compute the electric field via the R-W integral, the polar angle ($\theta_d$) defined with respect to the focus of the imaging lens is converted to the polar angle ($\theta$) defined with respect to the focus of the objective lens by using Snell's law and equation of magnification, $n_i \sin\theta = M n_a \sin\theta_d$, where $M$ is the effective magnification of a microscope.[3] For an imaging system with large $M$, $\sin\theta_d$ can be approximated to $\theta_d$ because the semi-aperture angle of the imaging lens in such a system is very small. Thus, the following angular relations are useful in the above equation: $d\theta_d = (n_i/M) \cos\theta \, d\theta$ and $\sin\theta_d \, d\theta_d = (n_i/M)^2 (\sin 2\theta / 2) d\theta$ where $n_a$ is set to be 1.[4] In our microscope setup, the magnification of the objective lens is 62 if the tube lens with a focal length of 180 mm is used as recommended. In fact, we used a tube lens with a focal length of 500 mm and thus, the effective magnification of our microscope is 170.

$$\vec{E}(r) = -\frac{ik}{2\pi} \int_0^{\theta_{max}} \int_0^{2\pi} \vec{E}_\infty(\theta_d, \phi) e^{ikz_d \cos\theta_d} e^{ikr\sin\theta_d \cos(\phi - \phi_d)} e^{ik\Lambda_{OPD}} \sqrt{\frac{n_i \cos\theta_d}{n_a \cos\theta}} \left(\frac{n_i}{M}\right)^2 \left(\frac{\sin 2\theta}{2}\right) d\phi d\theta \tag{S8}$$

where $\theta_{max}$ is the semi-aperture angle of $\theta$ and related to $\theta_{d,max}$ by $n_i \sin\theta_{max} = Mn_a \sin\theta_{d,max}$. To simplify this integral, the azimuthal angle ($\phi$) is integrated first by using the following relations:

$$\int_0^{2\pi} \cos n\phi \, e^{ix\cos(\phi-\varphi)} d\phi = 2\pi(i^n)J_n(x)\cos n\varphi \tag{S9}$$

$$\int_0^{2\pi} \sin n\phi \, e^{ix\cos(\phi-\varphi)} d\phi = 2\pi(i^n)J_n(x)\sin n\varphi \tag{S10}$$

The electric field is then simplified as follows:

$$\vec{E}(r) = -\frac{ik}{2}\left(\frac{n_i}{M}\right)^2 \int_0^{\theta_{max}} \vec{E}_a e^{ikz_d \cos\theta_d} e^{ik\Lambda_{OPD}} \sqrt{\frac{n_i \cos\theta_d}{n_a \cos\theta}} \sin 2\theta \, d\theta \tag{S11}$$

where the strength vectors $\vec{E}_a$ for reference and scattered fields are

$$\vec{E}_{a,\text{ref}} = \frac{1}{2}C_R \cdot i(r_p + r_s)\begin{bmatrix}(1-\cos\theta_d)J_2(kr\sin\theta_d)\sin 2\phi_d \\ (1+\cos\theta_d)J_0(kr\sin\theta_d)-(1-\cos\theta_d)J_2(kr\sin\theta_d)\cos 2\phi_d\end{bmatrix} \tag{S12}$$

$$\vec{E}_{a,\text{scat}} = \frac{1}{2}C_S \cdot i(S_2 t_p + S_1 t_s)\begin{bmatrix}(1-\cos\theta_d)J_2(kr\sin\theta_d)\sin 2\phi_d \\ (1+\cos\theta_d)J_0(kr\sin\theta_d)-(1-\cos\theta_d)J_2(kr\sin\theta_d)\cos 2\phi_d\end{bmatrix} \tag{S13}$$

where $\sin\theta_d = (1/M)n_i \sin\theta$ and $\cos\theta_d = \sqrt{1-(n_i \sin\theta/M)^2}$.

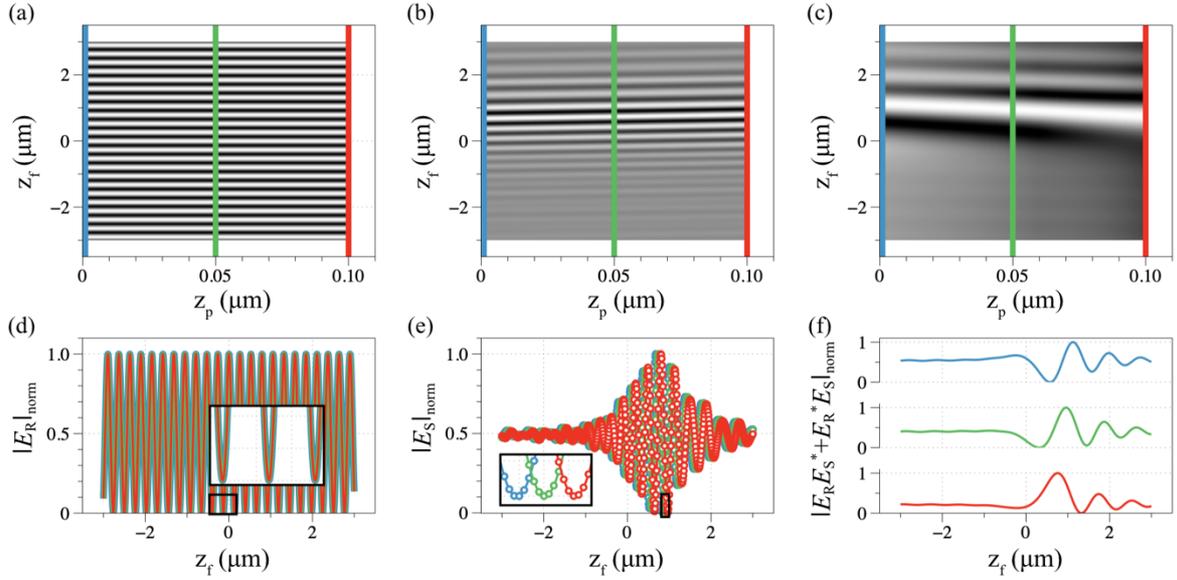

**Fig. S3** Stacked axial contrast profile as a function of the particle height ($z_p$). (a) Reference field ($|E_R|$), (b) scattered field ($|E_S|$) and (c) interference term ($|E_R E_S^* + E_R^* E_S|$) calculated in the range of $z_p = 0 \sim 100$ nm with $n_i = 1.51$. All calculated values are scaled to fit to [0,1] and displayed in a gray scale. The blue, green, and red colored lines indicate $z_p = 0$, 50, and 100 nm, respectively. (d-f) Cross-sections of the reference field (d), the scattered field (e), and the interference term (f). Insets in (d) and (e) show the phase variation of the fields and highlight the sensitivity of the fields to different values of $z_p$. The axial intensity profile by the interference term exhibits more pronounced variation with $z_p$ accounting for sensitive detection of the size of nano-objects by iSCAT. In this figure (a-f), $z_f$ is the computational input parameter before the offset ($\delta z$) is adjusted (see main text for details).

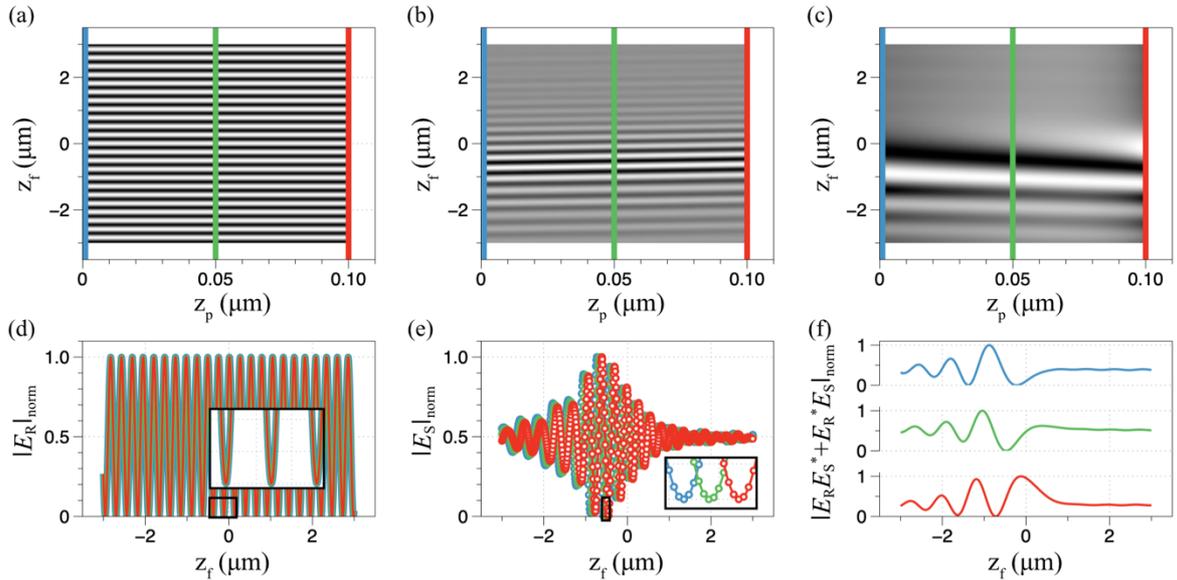

**Fig. S4** Simulations performed and results displayed identically to Fig. S3 except for $n_i = 1.53$.

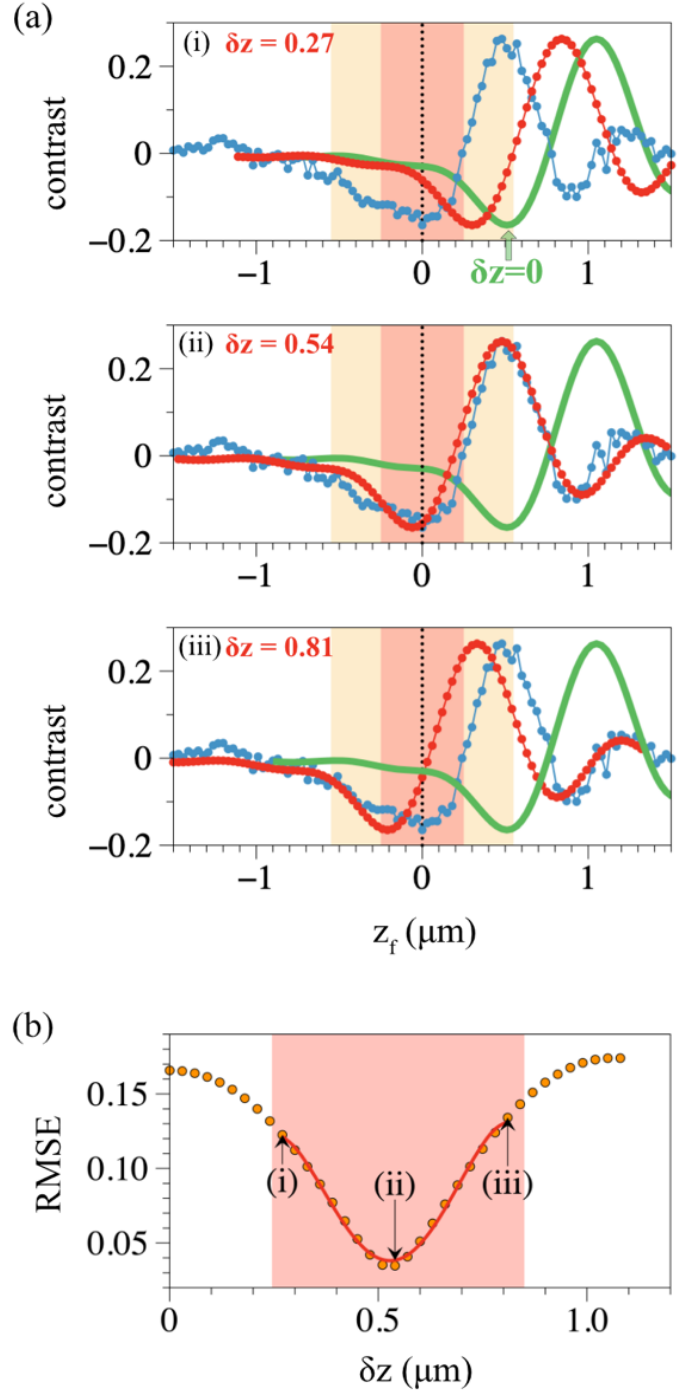

**Fig. S5** Root-mean-squared-error (RMSE) estimation to find the best theoretical axial intensity profile ($I_{sim}$) with a given set of $n_i$ and $z_p$ as a function of the relative focal position ($z$) that matches the experimental ($I_{exp}$) profile with the highest similarity. (a) Demonstration of the profile matching procedure via translation of the simulated profile ($I_{sim}$) along $z_f$ by $\delta z$. The RMSE value is computed from $I_{exp}$ (blue line with dots) and $I_{sim}$ (red line with dots) over the shaded region with good overlap. (b) RMSE values as a function of $\delta z$, which is the offset applied to $I_{sim}$. Once the value of $\delta z$ to yield the least RMSE is found, the offset is denoted as $\delta z^*$. (i) $\delta z$ = 0.27 μm, (ii) $\delta z$ = 0.54 μm = $\delta z^*$, and (iii) $\delta z$ = 0.81 μm with respect to the $\delta z$ = 0 μm (green solid line in (a)). We fitted the RMSE values to a 4$^{th}$ polynomial function to determine the representative RMSE for the given set of $n_i$ an $z_p$.

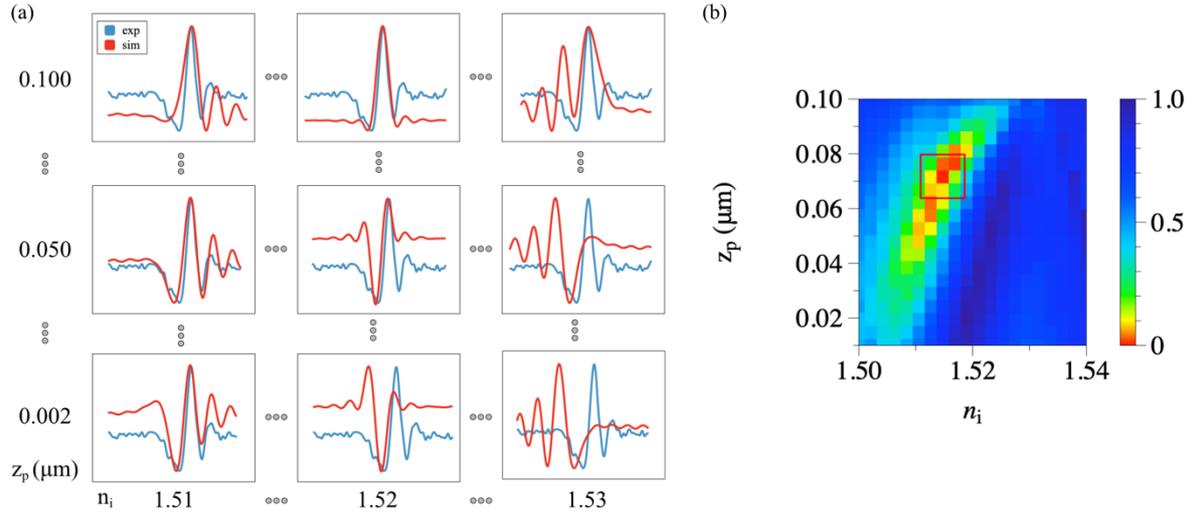

**Fig. S6** RMSE estimation to determine the model parameters, $n_i$ and $z_p$ for a 60-nm latex bead. (a) Illustration of the parameter-finding procedure by minimizing the RMSE between the measured axial contrast profile (blue line: experimental) and the computed axial contrast profile over the whole parameter space of $n_i$ and $z_p$ (red line: simulated). (b) Pseudo-color maps of RMSE re-scaled to fit to [0,1] for various $n_i$ and $z_p$. The minimum RMSE marked by a red box represents the best fit of the PSF modeling for the measured axial contrast profile. The $n_i$ and $z_p$ at the minimum RMSE are used as the parameters to generate the PSF with the highest similarity.

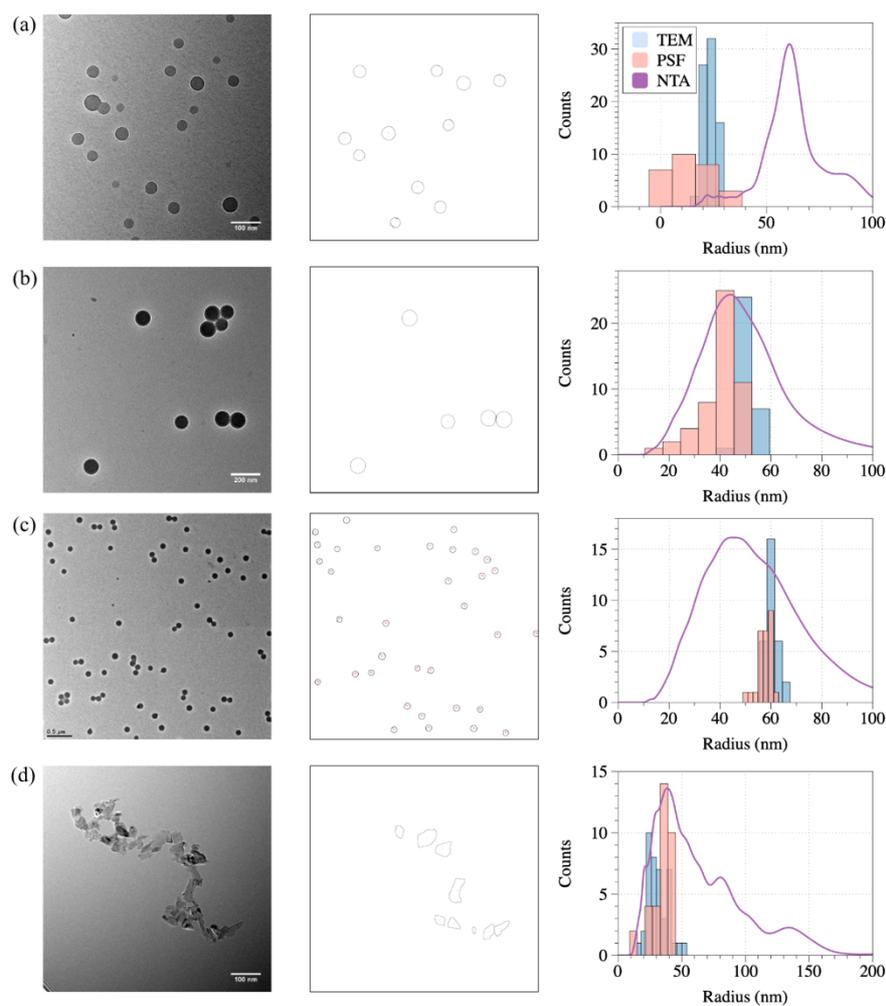

**Fig. S7** (First column) Representative TEM images that were analyzed to obtain the size of PS (a), latex ((b) and (c)) and fND (d) particles. (Second column) Outlines of detected particles used for calculation of the Feret diameter. (Third column) Histograms of size of particles measured by different techniques (TEM, PSF (iSCAT), and NTA analysis). Nominal radius ($R_{nom}$): (a) 20 nm PS beads, (b) 50 nm latex beads, (c) 60 nm latex beads, and (d) 40 nm for commercial fNDs. Radius deduced from TEM images was obtained by measuring the Feret diameter. Scale bar: (a) 100 nm, (b) 200 nm, (c) 500 nm, (d) 100 nm.

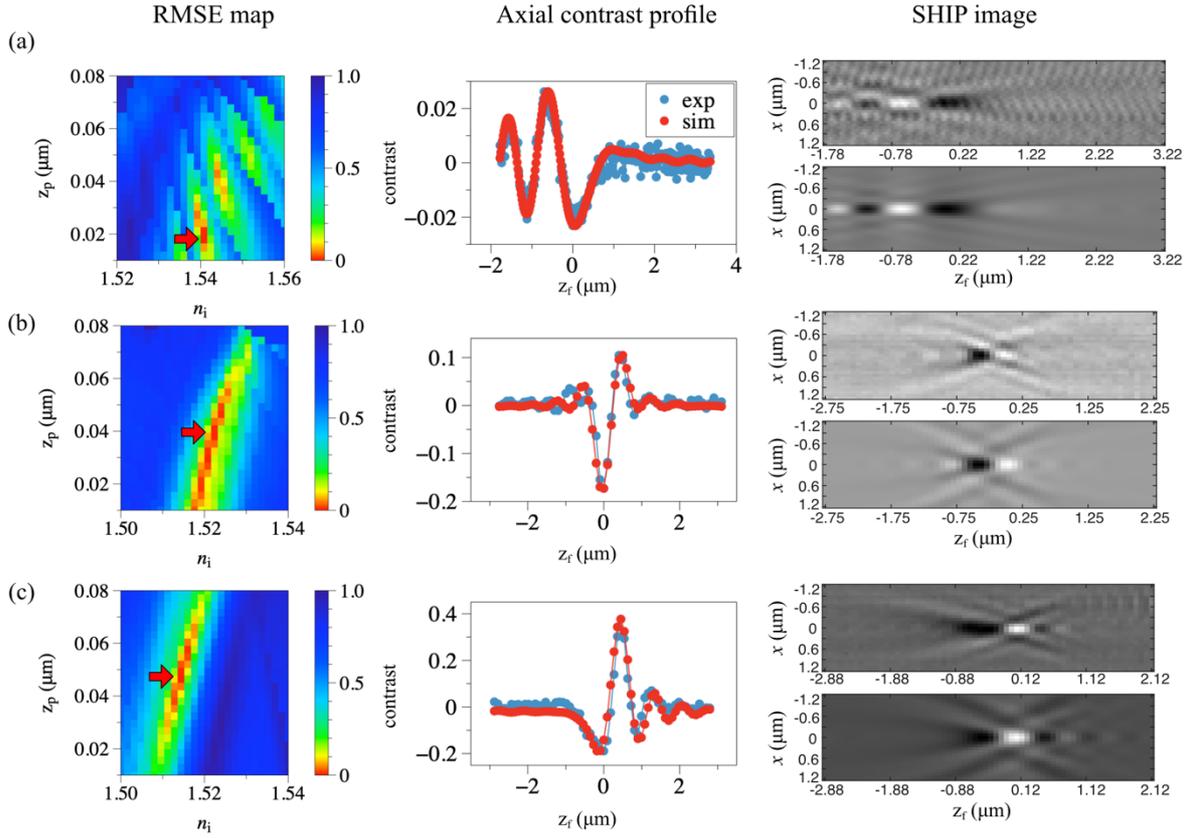

**Fig. S8** Representative results of PSF modeling analysis for different-sized dielectric nanoparticles ((a) $R_{nom}$ = 20 nm, (b) 50 nm, and (c) 60 nm). The RMSE map shows the representative RMSE values for the whole parameter space of $n_i$ and $z_p$, trying to find the best fit (marked by a red arrow). The axial contrast profile in the middle of each row shows the best simulation curve (red line with dots) for each experimental curve (blue line with dots). The SHIP images, one by experiment (top) and one by simulation (bottom) with the best parameter values: the fit results $(n_i, z_p)$ = (1.538, 0.022 μm) for (a), $(n_i, z_p)$ = (1.519, 0.039 μm) for (b), and $(n_i, z_p)$ = (1.515, 0.052 μm) for (c) with design parameters $(n_s, n_p, n_i^*, n_g^*, t_i^*, t_g^*)$ = (1.33, 1.59, 1.52, 1.52, 100 μm, 170 μm), exhibit clear similarity, indicating the reliability of the parameters obtained here.

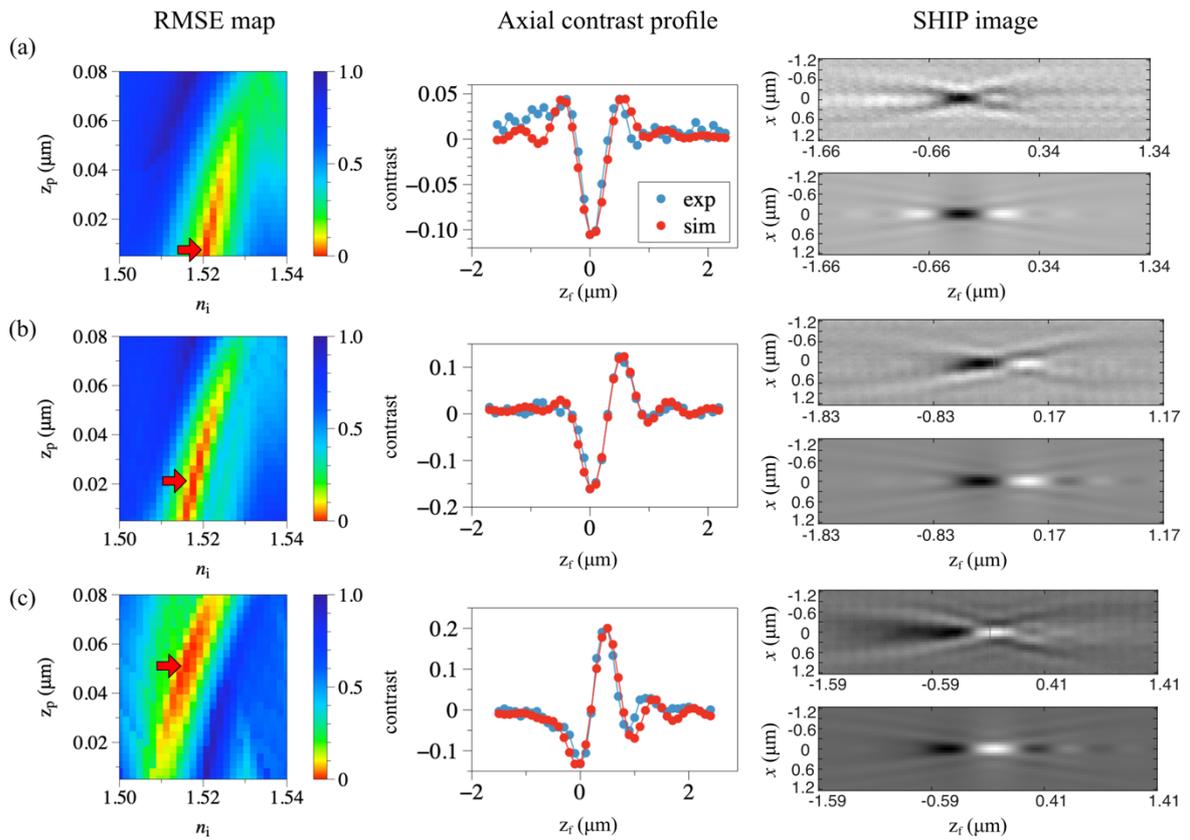

**Fig. S9** Representative results of PSF modeling analysis for different-sized fluorescent nanodiamonds (fND) particles from the same product number (900172, Sigma-Aldrich). This figure is presented in the same manner as Figure S8. The SHIP images, one by experiment (top) and one by simulation (bottom) with the best parameter values : the fit results $(n_i, z_p) = (1.521, 0.012\ \mu m)$ for (a), $(n_i, z_p) = (1.518, 0.022\ \mu m)$ for (b), and $(n_i, z_p) = (1.516, 0.045\ \mu m)$ for (c) with design parameters $(n_s, n_p, n_i^*, n_g^*, t_i^*, t_g^*) = (1.33, 2.4, 1.52, 1.52, 100\ \mu m, 170\ \mu m)$, exhibit clear similarity, indicating the reliability of the parameters obtained here. It is also notable that the size of fND is highly heterogeneous.